\documentclass[conference,letter]{IEEEtran}
\IEEEoverridecommandlockouts

\usepackage{graphicx}
\usepackage{balance}
\usepackage[T1]{fontenc}
\usepackage{ifthen}
\usepackage[cmex10]{amsmath} %

\usepackage[lined,boxed,commentsnumbered,linesnumbered, ruled]{algorithm2e}
\usepackage[url,hyperrefblack,notheorems,IEEEtran]{research17}
\usepackage{comment}
\usepackage{booktabs} %
\usepackage{mathtools}
\usepackage{amsmath,amssymb,amsfonts,amsthm}
\usepackage{tikz} %
\tikzset{font={\fontsize{10pt}{12}\selectfont}}
\usetikzlibrary{shapes, patterns,decorations.text, decorations.pathreplacing,plotmarks}

\newtheorem{lemma}{\mylemmaname}
\newtheorem{theorem}{\mytheoremname}
\newtheorem{definition}{\mydefinitionname}
\newtheorem{proposition}{\mypropositionname}

\newtheorem{example}{\myexamplename}

\usepackage[capitalize]{cleveref}

\crefname{equation}{\unskip}{\unskip}
\crefname{claim}{Claim}{Claims} %
\usepackage{array}
\usepackage{multirow}
\newcolumntype{C}[1]{>{\centering\arraybackslash}p{#1}}
\allowdisplaybreaks

\renewcommand{\deg}[1]{\operatorname{deg}\left(#1\right)} %
\renewcommand{\vect}[1]{\vectg{#1}} %
\renewcommand{\vmat}[1]{\bm{\mat{#1}}} %
\newcommand{\code}[1]{\mathscr{#1}} %
\newcommand*{\Resize}[2][4]{\resizebox{#1}{!}{\ensuremath{#2}}} %
\makeatletter
\renewcommand*\env@matrix[1][*\c@MaxMatrixCols c]{%
  \hskip -\arraycolsep
  \let\@ifnextchar\new@ifnextchar
  \array{#1}}
\makeatother
\newcommand{\HP}[1]{\HH\left(#1\right)} 
\newcommand{\eHP}[1]{\HH(#1)}

\newcommand{\HPcond}[2]{\HH\left(#1 \kern0.1em\middle|\kern0.1em #2\right)}
\newcommand{\eHPcond}[2]{\HH(#1 \kern0.1em|\kern0.1em #2)} 
\newcommand{\bigHPcond}[2]{\HH\bigl(#1 \kern-0.1em \bigm| \kern-0.1em#2\bigr)}
\newcommand{\BigHPcond}[2]{\HH\Bigl(#1 \kern-0.1em \Bigm| \kern-0.1em#2\Bigr)}
\newcommand{\MI}[2]{\II\left(#1 \kern0.1em{;}\kern0.1em #2\right)} 
\newcommand{\eMI}[2]{\II(#1 \kern0.1em{;}\kern0.1em #2)} 
\newcommand{\bigMI}[2]{\II\bigl(#1 \kern0.1em{;}\kern0.1em #2\bigr)}
\newcommand{\BigMI}[2]{\II\Bigl(#1 \kern0.1em{;}\kern0.1em #2\Bigr)}
\newcommand{\MIcond}[3]{\II\left(#1 \kern0.1em{;}\kern0.1em #2 \kern0.1em\middle|\kern0.1em #3\right)}
\newcommand{\eMIcond}[3]{\II(#1 \kern0.1em{;}\kern0.1em #2 \kern0.1em|\kern0.1em #3)} 
\newcommand{\bigMIcond}[3]{\II\bigl(#1 \kern0.1em{;}\kern0.1em #2 \kern-0.1em \bigm| \kern-0.1em#3\bigr)}
\newcommand{\BigMIcond}[3]{\II\Bigl(#1 \kern0.1em{;}\kern0.1em #2 \kern-0.1em \Bigm| \kern-0.1em#3\Bigr)}
\renewcommand{\r}{\color{red}} %
\renewcommand{\b}{\color{blue}} %
\DeclareSymbolFont{matha}{OML}{txmi}{m}{it}%
\DeclareMathSymbol{\varv}{\mathord}{matha}{118}
\usepackage{upgreek}

\begin{document}
\sloppy \title{Private Polynomial Computation for Noncolluding Coded Databases\thanks{This work is supported by US NSF grant CNS-1526547 and the Research Council of Norway (grant 240985/F20).}}

\author{\IEEEauthorblockN{Sarah A.~Obead\IEEEauthorrefmark{2}, Hsuan-Yin Lin\IEEEauthorrefmark{3}, Eirik
    Rosnes\IEEEauthorrefmark{3}, and J{\"o}rg Kliewer\IEEEauthorrefmark{2} \\
    \IEEEauthorblockA{\IEEEauthorrefmark{2}Helen and John C.~Hartmann
      Department of Electrical and Computer Engineering \\ New Jersey Institute of Technology, Newark, New Jersey 07102, USA\\}
    \IEEEauthorblockA{\IEEEauthorrefmark{3}Simula UiB, N--5008 Bergen, Norway}}}

\maketitle

\begin{abstract}
  We consider private polynomial computation (PPC) over noncolluding coded
  databases. In such a setting a user wishes to compute a multivariate
  polynomial of degree at most $g$ over $f$ variables (or messages) stored
  in multiple databases while revealing no information about the desired
  polynomial to the databases. We construct two novel PPC
  schemes, where the first is a generalization of our previous work in
  private linear computation for coded databases. In this scheme we consider
  Reed-Solomon coded databases with Lagrange encoding, which leverages ideas
  from recently proposed star-product private information retrieval and
  Lagrange coded computation. The second scheme considers the special case
  of coded databases with systematic Lagrange encoding. Both schemes yield 
  improved  rates compared to the best known schemes from the literature for
  a small number of messages, while in the asymptotic case the rates
  match. 
\end{abstract}

\vspace{-0.8ex}
\section{Introduction}
\label{sec:introduction}
\vspace{-0.5ex}
The notion of private information retrieval (PIR) was introduced by Chor \emph{et al.} in the computer science community
\cite{ChorKushilevitzGoldreichSudan95_1}. The goal of PIR is to allow a user to privately access an arbitrary message
stored in a set of databases, i.e., without revealing any information of the identity of the requested message to each
database.
The design of PIR protocols has focused on the case when multiple databases store the messages. This connects to the
active and renowned research area of distributed storage systems (DSSs), where the messages are encoded by an $[n,k]$
linear code and then distributed and stored across $n$ storage nodes.
The study and design of efficient PIR protocols for coded DSSs have attracted a great deal of attention in recent years
\cite{SunJafar17_1,FreijHollantiGnilkeHollantiKarpuk17_1,BanawanUlukus18_1,TajeddineGnilkeElRouayheb18_1,KumarLinRosnesGraellAmat19_1app}.

Private computation is a generalization of PIR that addresses the private computation for functions of the stored
messages
\cite{MirmohseniMaddahAli18_1,SunJafar19_1app,ChenWangJafar18_1,Karpuk18_1,ObeadKliewer18_1,ObeadLinRosnesKliewer18_1,RavivKarpuk18_1sub}. The
scenario of noncolluding replicated databases for linear functions is considered in
\cite{MirmohseniMaddahAli18_1,SunJafar19_1app} and referred to as private linear computation (PLC).
The coded case is addressed in \cite{ObeadKliewer18_1,Karpuk18_1,ObeadLinRosnesKliewer18_1,RavivKarpuk18_1sub}. In
particular, in \cite{ObeadKliewer18_1,ObeadLinRosnesKliewer18_1} we proposed a PLC scheme based on maximum distance
separable (MDS) coded storage, where the obtained PLC capacity is equal to the MDS-coded PIR capacity in
\cite{BanawanUlukus18_1}. %
In \cite{Karpuk18_1}, private polynomial computation (PPC) over $t$ colluding and systematically coded databases is
considered by generalizing the star-product PIR scheme of \cite{FreijHollantiGnilkeHollantiKarpuk17_1}. In that work,
functions are computed that are polynomials of degree at most $g$, and a private computation rate equal to the best
asymptotic PIR rate (when the number of messages tends to infinity) of MDS-coded storage is achieved for $g=t=1$.  An
alternative PPC approach was recently proposed in \cite{RavivKarpuk18_1sub} by employing Reed-Solomon (RS) coded
databases with Lagrange encoding. For low code rates, the scheme improves on the private computation rate of
\cite{Karpuk18_1}.

In this work, we present two new approaches for PPC over coded databases by leveraging our previous works for PLC in
\cite{ObeadKliewer18_1}, \cite{ObeadLinRosnesKliewer18_1}, ideas from star-product PIR
  \cite{FreijHollantiGnilkeHollantiKarpuk17_1}, and Lagrange coded computation
  \cite{YuLiRavivKalanSoltanolkotabiAvestimehr19_1}.  Our schemes apply to noncolluding RS-coded databases with
Lagrange encoding.
Compared to the scheme in \cite{RavivKarpuk18_1sub}, our first proposed PPC scheme yields a higher private computation
rate when the number of messages is small. In addition, we construct a second PPC scheme for RS-coded databases with
systematic Lagrange encoding that improves on the rate of the PPC scheme presented in \cite{Karpuk18_1}. In both cases,
as the number of messages tends to infinity, the rate approaches those of \cite{RavivKarpuk18_1sub} and
\cite{Karpuk18_1}, respectively. For the outer bound, we adopt our
  coded PLC capacity of \cite[Thm.~2]{ObeadLinRosnesKliewer18_1} since PPC can be seen as an extension of PLC.

\section{Definitions and Problem Statement}
\label{sec:definitions-problem}

\subsection{Notation}
\label{sec:notation}
\vspace{-0.5ex}
We denote by $\Naturals$ the set of all positive integers, $[a]\eqdef\{1,2,\ldots,a\}$, and
$[a:b]\eqdef\{a,a+1,\ldots,b\}$ for $a,b\in \Naturals$, $a \leq b$.  
A random variable is denoted by a capital Roman letter, e.g., $X$, while its realization is denoted by the corresponding small
Roman letter, e.g., $x$. Vectors are boldfaced, e.g., $\vect{X}$ denotes a random vector and $\vect{x}$ denotes a
deterministic vector, respectively. Random matrices are represented by bold sans serif letters, e.g., $\vmat{X}$, where $\mat{X}$
represents its realization. In addition, sets are denoted by calligraphic uppercase letters, e.g., $\set{X}$. %
$\trans{(\cdot)}$ denotes the transpose operator, 
$\HP{X}$ represents the entropy of $X$, and $\MI{X}{Y}$ the mutual
information between $X$ and $Y$. The binomial coefficient of $a$ over $b$, $a,b \in \{0\}\cup \Naturals$, is denoted by $a \choose b$ where  ${a \choose b} \eqdef 0$ if $a < b$. We use the customary code parameters $[n,k]$ to denote a code $\code{C}$
over the finite field $\Field_q$ of blocklength $n$ and dimension $k$.
The function $\chi(\vect{x})$ denotes the support of a vector $\vect{x}$, and the linear span of a set of vectors $\{\vect{x}_1,\ldots,\vect{x}_a\}$, $a\in \Naturals$, is denoted by $\mathsf{span}\{\vect{x}_1,\ldots,\vect{x}_a\}$. 
A monomial ${\vect{W}}^{\vect{i}}$ in $f$ variables $W^{(1)},\ldots,W^{(f)}$ with degree $g$ is written as
${\vect{W}}^{\vect{i}} = (W^{(1)})^{i_1}(W^{(2)})^{i_2}\cdots(W^{(f)})^{i_f}$, where
$\vect{i}\eqdef(i_1,\ldots,i_f)\in (\{0\} \cup \Naturals)^{f}$ is the exponent vector with $\mathsf{wt}(\vect{i})\eqdef\sum_{j=1}^{f}i_j = g$. 
Finally, a polynomial $\phi(\vect{W})$ of degree at
most $g$ is  represented as $\phi(\vect{W})=\sum_{\vect{i}:\textsf{wt}(\vect{i}) \leq g} a_{\vect{i}}{\vect{W}}^{\vect{i}}$, $a_{\vect{i}} \in \Field_q$. $\Field_q[z]$ denotes the set of all univariate polynomials over  $\Field_q$ in the variable $z$. We denote by $\deg{\phi(z)}$ the degree %
of a polynomial $\phi(z) \in \Field_q[z]$.

\subsection{Preliminaries}
\label{sec:preliminaries}

\begin{definition}[Star-product]
  Let $\code{C}$ and $\code{D}$ be two linear codes of length $n$ over
  $\Field_q$.   The star-product (Hadamard product) of 
  $\vect{v}=(v_1,\ldots,v_n)  \in\code{C}$ and $\vect{u} = (u_1,\ldots,u_n) \in\code{D}$ is defined as
  $\vect{v}\star\vect{u}=(v_1u_1,\ldots,v_nu_n)\in \Field_q^n$.
  Further, the star-product of $\code{C}$ and $\code{D}$, denoted by $\code{C}\star\code{D}$, is defined by
  $\mathsf{span}\{\vect{v}\star\vect{u}:\vect{v}\in\code{C}, \vect{u}\in\code {D}\}$
  and the $g$-fold star-product of  $\code{C}$ with itself is given by
  $\code{C}^{\star g}=\mathsf{span}\{\vect{v}_1\star\cdots\star\vect{v}_g:\vect{v}_i\in\code{C}, i\in[g]\}.$
\end{definition}

\begin{definition}[Reed-Solomon code]
  Let $\vect{\alpha}=(\alpha_1,\ldots,\alpha_n)$ be a vector of $n$ distinct nonzero elements of $\Field_q$. For $n\in\Naturals$, $k \in [n]$, and
  $q > n$, 
  the $[n,k]$ RS code (over $\Field_q$) is defined as
  \begin{IEEEeqnarray}{rCl}
    \set{RS}_{k}(\vect{\alpha})& \eqdef & \{(\phi(\alpha_1),\ldots,\phi(\alpha_n))\colon \phi\in\Field_q[z],\, \deg{\phi}<k \}.
    \IEEEeqnarraynumspace\label{eq:RScodeDef}
  \end{IEEEeqnarray}
\end{definition}

It is well-known that RS codes are MDS codes that behave well under the star-product. %
We state the following proposition that was introduced in
\cite{FreijHollantiGnilkeHollantiKarpuk17_1}.
	  
\begin{proposition}\label{prop:RSstarProduct}
  Let ${\cal RS}_{k}(\vect{\alpha})$ be a length-$n$ RS code. Then, for $g\in\Naturals$, the $g$-fold star-product of
  ${\cal RS}_{k}(\vect{\alpha})$ with itself is the RS code given by
  ${\cal RS}_k^{\star g}(\vect{\alpha})={\cal RS}_{\min{\{g(k-1)+1,n\}}}(\vect{\alpha})$.
\end{proposition}

  Let $\vect{\gamma}=(\gamma_1,\ldots,\gamma_k)$ be a vector of $k$ distinct elements of $\Field_q$. For a message vector
  $\vect{W}=(W_1,\ldots,W_k)$, let $\ell(z)\in \Field_q[z]$ be a polynomial of degree at most $k-1$ such that
  $\ell(\gamma_i)=W_{i}$ for all $i\in[k]$. Using the Lagrange interpolation formula we present this polynomial as
  $\ell(z)=\sum_{i\in[k]} W_i \iota_i(z)$, where $\iota_i(z)$ is the Lagrange basis polynomial 
  \vspace{-0.4ex}
  \begin{IEEEeqnarray*}{rCl}
    \iota_{i}(z)=\prod_{t\in[k]\setminus \{i\}}\frac{z-\gamma_t}{\gamma_i-\gamma_t}.
  \end{IEEEeqnarray*}

  It has been shown in \cite{RavivKarpuk18_1sub} that Lagrange encoding is equivalent to the choice of a specific basis
  for an RS code. Thus, for encoding we choose the set of Lagrange basis polynomials as the code generating polynomials
  of \eqref{eq:RScodeDef} \cite{YuLiRavivKalanSoltanolkotabiAvestimehr19_1}. Thus, a generator matrix of
  ${\cal RS}_{k}(\vect{\alpha})$ is
  $\mat{G}_{\mathcal{RS}_{k}}(\vect{\alpha,\gamma}) = (\iota_{i}(\alpha_j))$, $i\in[k]$, $j\in[n]$.
  Note that if we choose $\gamma_i=\alpha_i$ for $i\in[k]$, then the generator matrix
  $\mat{G}_{\mathcal{RS}_{k}}(\vect{\alpha,\gamma})$ becomes systematic.

The set $\{\vect{W}^{\vect{i}}: \vect{i} \in (\{ 0  \} \cup \Naturals)^f,\, 1\leq \textsf{wt}(\vect{i})\leq g\}$ of all monomials in $f$ variables of degree at most $g$ has size
\begin{IEEEeqnarray*}{rCl}
  \const{M}(f,g)\eqdef\sum_{h=1}^{g}{h+f-1\choose h} = {g+f\choose g}-1,
\end{IEEEeqnarray*}
and the total number of polynomials in $f$ variables of degree at most $g$ generated with all possible distinct (up to scalar multiplication) $\const{M}(f,g)$-dimensional coefficients
vectors defined over $\Field_q$ is equal to $\mu(f,g) \eqdef \frac{q^{\const{M}(f,g)}-1}{q-1}$.

\subsection{System Model}
\label{sec:system-model}

An RS-coded DSS is described as follows. The DSS stores in total $f$ independent messages
$\vmat{W}^{(1)},\ldots,\vmat{W}^{(f)}$, where each message $\vmat{W}^{(m)}=\bigl(W_{i,j}^{(m)}\bigr)$, $m\in [f]$, is
a random $\beta\times k$ matrix with some $\beta,k \in\Naturals$, where each entry is chosen independently and uniformly
at random from $\Field_q$. Thus, $\eHP{\vmat{W}^{(m)}}=\beta k\eqdef\const{L},\,\forall\,m\in[f]$ (in $q$-ary units).

Each message is encoded using an $[n,k]$ RS code as follows. Let
$\vect{W}^{(m)}_i=\bigl(W^{(m)}_{i,1},\ldots,W^{(m)}_{i,k}\bigr)$, $i\in [\beta]$, be a message vector corresponding
to the $i$-th row of $\vmat{W}^{(m)}$. Each $\vect{W}^{(m)}_i$ is encoded by an RS code ${\cal RS}_{k}(\vect{\alpha})$
with evaluation vector $\vect{\alpha}=(\alpha_1,\ldots,\alpha_n)$ over $\Field_q$ into a length-$n$ codeword
$\vect{C}^{(m)}_i$ where
$\vect{C}^{(m)}_i= \vect{W}^{(m)}_i \mat{G}_{\mathcal{RS}_k}(\vect{\alpha},\vect{\gamma})=
\bigl(C^{(m)}_{i,1},\ldots,C^{(m)}_{i,n}\bigr)$ 
and $C^{(m)}_{i,j} = \ell_i^{(m)}(\alpha_j)$, $j\in[n]$,  where $\ell_i^{(m)}(z)$ is the Lagrange interpolation polynomial associated with the length-$k$ message segment
$\vect{W}^{(m)}_i$. The $\beta f$ generated codewords $\vect{C}_i^{(m)}$ are then arranged in the array
$\vmat{C}=\trans{\bigl(\trans{(\vmat{C}^{(1)})}|\ldots|\trans{(\vmat{C}^{(f)})}\bigr)}$ of dimensions
$\beta f \times n$, where
$\vmat{C}^{(m)}=\trans{\bigl(\trans{(\vect{C}^{(m)}_1)}|\ldots|\trans{(\vect{C}^{(m)}_{\beta})}\bigr)}$. The code
symbols $C_{1,j}^{(m)},\ldots,C_{\beta,j}^{(m)}$, $m\in[f]$, for all $f$ messages are stored on the $j$-th database,
$j\in[n]$.

\subsection{Private Polynomial Computation for RS-Coded DSSs}
\label{sec:PPC_RScoded-DSSs}

We consider the case of $n$ noncolluding databases. A user wishes to
privately compute exactly one polynomial out of $\mu$ \emph{candidate} polynomial functions
$\vmat{X}^{(1)},\ldots,\vmat{X}^{(\mu)}$ from  the RS-coded DSS while
keeping the requested index private from each database. The polynomial function
$\vmat{X}^{(v)}=\bigl(\phi^{(v)}(\vect{W}_{i,j})\bigr)$, where $\vect{W}_{i,j} = (W_{i,j}^{(1)},\ldots,W_{i,j}^{(f)})$, is a $\beta\times k$ random matrix for some polynomial $\phi^{(v)}$, where each $\phi^{(v)}(\vect{W}_{i,j}) \in \Field_q$ is independent and distributed according to some probability mass function~$P_{X_v}$. %
Thus, $\eHP{\vmat{X}^{(v)}}=\const{L}\eHP{X_v},\,\forall\,v\in[\mu]$, and
$\eHP{\vmat{X}^{(1)},\ldots,\vmat{X}^{(\mu)}}=\const{L}\HP{X_1,\ldots,X_{\mu}}$.

  Consider an RS-coded DSS with $n$ noncolluding databases storing $f$ messages. The user wishes to retrieve the $v$-th
  polynomial function $\vmat{X}^{(v)}$, $v\in[\mu]$, from the available information from queries $Q^{(v)}_j$ and answer
  strings $A^{(v)}_j$, $j\in[n]$. For a PPC protocol, the following conditions must be satisfied $\forall\,v\in[\mu]$,
  \begin{IEEEeqnarray*}{rCl}    
    \IEEEeqnarraymulticol{3}{l}{%
      \text{[Privacy]} }\nonumber\\*\qquad\qquad%
    && \bigMI{v}{Q^{(v)}_j,A^{(v)}_j,\vmat{X}^{(1)},\ldots,\vmat{X}^{(\mu)}}=0,\,\forall\,j\in[n],
    \\
    \IEEEeqnarraymulticol{3}{l}{%
      \text{[Recovery]} }\nonumber\\*\qquad\qquad%
    && \bigHPcond{\vmat{X}^{(v)}}{A^{(v)}_1,\ldots,A^{(v)}_n,Q^{(v)}_1,\ldots,Q^{(v)}_n}=0.
  \end{IEEEeqnarray*}

\begin{definition}[PPC rate for RS-coded DSSs]
  \label{def:def_PPCrate}
  The rate of a PPC scheme, denoted by $\const{R}$, is defined as $\const{R}=\const{L}/\const{D}$, where $\const{D}$ is
  the total required download cost.\footnote{In order to compare with the PPC schemes from
    \cite{Karpuk18_1,RavivKarpuk18_1sub}, we use a slightly imprecise definition of the PPC rate. The exact
    information-theoretic PPC rate is  defined as the ratio of the minimum desired polynomial function size
    $\const{L}\min_{v \in [\mu]} \HP{X_v}$ over the total required download cost $\const{D}$. }
\end{definition}

\begin{definition}[$\tau$-sum]
For $\tau\in[\mu]$, a sum $\phi^{(v_1)}(\vect{C}_{i_1,j}) + \cdots + \phi^{(v_\tau)}(\vect{C}_{i_\tau,j})$, where $\vect{C}_{i,j} = (C^{(1)}_{i,j},\ldots,C^{(f)}_{i,j})$,  $i \in [\beta]$, $j\in[n]$, of
$\tau$ distinct candidate polynomial function evaluations is called a $\tau$-sum for any $(i_1,\ldots,i_{\tau})\in [\beta]^\tau$, and
$\{v_1,\ldots,v_{\tau}\}\subseteq [\mu]$ determines the type of the $\tau$-sum.
\end{definition}

\section{A General PPC Scheme for RS-Coded DSSs With Lagrange Encoding}
\label{sec:achievable-scheme_PMC}
In the following we build a PPC scheme based on Lagrange
encoding and our PLC scheme
in \cite{ObeadLinRosnesKliewer18_1}. Note that a polynomial can be written as
a linear combination of monomials, and therefore any private monomial
computation (PMC) scheme is a special case of PPC. Thus, a PPC scheme can be
obtained from a PLC scheme by replacing independent messages with a monomial basis. We first
discuss the PPC case in general and then provide an example for the
special case of PMC.

\subsection{Lagrange Coded Computation}
\label{sec:lagrange-computation}
\vspace{-0.5ex}
Lagrange coded computation \cite{YuLiRavivKalanSoltanolkotabiAvestimehr19_1} is a framework that can be applied to any
function computation when the function of interest is a multivariate polynomial of the messages. We extend the
application of this framework to PMC and PPC by utilizing the following argument.

Recall that $\ell_t^{(m)}(z)$, $t\in[\beta]$, $m\in[f]$, 
evaluated at $\gamma_j$ results in an
information symbol $W^{(m)}_{t,j}$ and when evaluated at $\alpha_j$ we
obtain a code symbol $C^{(m)}_{t,j}$. Let
$\vect{\ell}_t(z)=(\ell_t^{(1)}(z), \ldots, \ell_t^{(f)}(z))$ be a vector of $f$ Lagrange interpolation polynomials
associated with the messages $\vect{W}^{(1)}_t,\ldots,\vect{W}^{(f)}_t$. Now, given a multivariate polynomial function
$\phi(\vect{W}_{t,j})$ of degree at most $g$, we introduce the composition function
$ \psi_t(z) =\phi(\vect{\ell}_t(z))$. Accordingly, evaluating $\psi_t(z)$ at any
$\gamma_j$, $j\in[k]$, is equal to evaluating the polynomial function over the uncoded information symbols, i.e., 
$\phi(\vect{W}_{t,j})$ and similarly, evaluating $\psi_t(z)$ at $\alpha_j$, $j \in [n]$, will result in the evaluation
of the polynomial function over the coded symbols, i.e.,  $\phi(\vect{C}_{t,j})$.  Since each Lagrange interpolation
polynomial of $\vect{\ell}_t(z)$ is a polynomial of degree at most $k-1$, it follows that $\deg{\psi_t(z)}\leq g(k-1)$ and we require up to 
$g(k-1)+1$ coefficients to interpolate and determine the polynomial $\psi_t(z)$. 

Note that $\psi_t(z)$ is a linear combination of monomials $z^{i}\in\Field_q[z]$, $i\leq g(k-1)$, and the underlying code  $\tilde{\code{C}}$ for  $(\psi_t(\alpha_1),\ldots,\psi_t(\alpha_n))$,  referred to as the \emph{decoding code},  is given by the
$g$-fold star-product ${\cal RS}_k^{\star g}(\vect{\alpha})$ of the storage code  
${\cal RS}_k(\vect{\alpha})$ according to \cite[Lem.~7]{RavivKarpuk18_1sub}. This is due to the fact that
the span
 of ${\cal RS}_k^{\star g}(\vect{\alpha})$
is given by  linear
combinations of codewords in ${\cal RS}_k^{\star g}(\vect{\alpha})$ where each code symbol represents a monomial. With other words,
to construct coded PPC schemes that retrieve polynomials of degree at most $g$, we require $g(k-1)+1 \leq n$ and $d_{\mathsf{min}}^{\tilde{\code{C}}} \geq n-(g(k-1)+1)+1$, where $d_{\mathsf{min}}^{\tilde{\code{C}}}$ denotes the minimum distance of $\tilde{\code{C}}$, to be able to decode the computation
correctly. It follows from Proposition~\ref{prop:RSstarProduct} that $\tilde{\code{C}} = {\cal RS}_{{\tilde{k}}}(\vect{\alpha})$ with dimension $\tilde{k} = \min\{g(k-1)+1,n\} = g(k-1)+1$ and $d_{\mathsf{min}}^{\tilde{\code{C}}} = n-\tilde{k}+1 = n-(g(k-1)+1)+1$. %

\subsection{PPC Achievable Rate Matrix}
\label{sec:ppc-achievable-rate}

Similar to \cite[Def.~3]{ObeadLinRosnesKliewer18_1},
where we introduce the notion of a PIR achievable rate matrix for the coded
PLC problem, we provide the following definition for the PPC case.
\begin{definition}
  \label{def:codes_CstarG-nonSysPPCachievable}
  A $\nu\times n$ binary matrix ${\mat{\Lambda}}_{\kappa,\nu}$ is called a \emph{PPC achievable rate matrix} for
  $(\code{C},\tilde{\code{C}})$ if the following conditions are satisfied.
  \begin{enumerate}
  \item \label{item:1} The Hamming weight of each column of ${\mat{\Lambda}}_{\kappa,\nu}$ is $\kappa$, and
  \item \label{item:2} for each matrix row $\vect{\lambda}_i$, $i\in [\nu]$,
    $\chi(\vect{\lambda}_i)$ is always an
    information set for $\tilde{\code{C}}$.
  \end{enumerate}
\end{definition}

\subsection{Redundancy Elimination}
\label{sec:PMC_Redundancy}

Here, we generalize the coded PLC scheme of \cite{ObeadLinRosnesKliewer18_1} in terms of exploiting the dependency
between the virtual messages. Since any polynomial is a linear function of the monomial basis of size $\const{M}(f,g)$,
a PPC scheme can be seen as a PLC scheme performed over a set of $\const{M}(f,g)$ messages. Hence, the
redundancy resulting from the linear dependencies between the virtual messages is also present for PPC and we can extend
\cite[Lem.~1]{ObeadLinRosnesKliewer18_1} and \cite[Lem.~1]{SunJafar19_1app} to our scheme. To exploit the dependency
between the virtual messages we adopt a similar sign assignment process to each queried symbol of the virtual monomial messages, based on the desired function index $v$ as introduced in \cite[Sec.~IV.B]{SunJafar19_1app}. This will
result in a uniquely solvable equation system from the different $\tau$-sum types 
given the side information available
from all other databases. By obtaining such a system of equations in each round $\tau\in[\mu]$ of the protocol, the user can determine some of the
answers offline.

Now, consider $1$-sum types, where we download individual segments of each virtual message including $f$
independent messages. For these types, the user can determine any polynomial
from the $f$ obtained  message segments. Based on this insight we can state the following
lemma.

\begin{lemma}
  \label{lem:redundancy}
  Let $\mu \in [f:\mu(f,g)]$ be the number of candidate polynomials, including the $f$ independent messages. For each query set, for all $v\in[\mu]$, each database $j\in[n]$, and based on the queried segments from the $f$ independent messages, there are $\mu-f\choose 1$ redundant $1$-sum types out of all possible types $\mu\choose 1$. On the other hand, for $\tau\in[2:\mu]$, there are $\max\{\mu-{\const M}(f,g), 0\} \choose \tau$ redundant $\tau$-sum types out of
  $\mu \choose \tau$ types. The number of nonredundant $\tau$-sum types with $\tau>1$ is given by 
    $\rho(\mu,\tau)\triangleq{\mu \choose \tau}-{\max\{\mu-{\const M}(f,g), 0\}\choose \tau}$.
\end{lemma}

\subsection{Achievable PPC Rate}
\label{sec:PPCrate_codes}

Since $\tilde{\code{C}}$ is an $[n,\tilde{k}]$ MDS code ($\code{C}$ is an RS code), there always exists a PPC achievable rate matrix
$\Lambda_{\kappa,\nu}$ with $(\kappa,\nu)=(\tilde{k},n)$ for $(\code{C},\tilde{\code{C}})$. Hence, using \cref{lem:redundancy} we can prove %
the following theorem.

\begin{theorem}
  \label{thm:PMCrate_LagrangeCoded-DSS}
  Consider a DSS that uses an $[n,k]$ RS code $\code{C}$ to store $f$ messages over $n$
  noncolluding databases using Lagrange encoding. Let $\mu \in [f:\mu(f,g)]$
  be the number of candidate polynomials to be computed of degree at most
  $g$, $g(k-1)+1 \leq n$,  including the $f$ independent messages.
  Then, the PPC rate 
  \begin{IEEEeqnarray*}{c}
    \const{R}_{\textnormal{PPC}}=\frac{k n^{\mu-1}}
    {f  {\tilde{k}}^{\mu}+\sum_{\tau=2}^{\mu} {\rho(\mu,\tau) }
      {\tilde{k}}^{\mu-\tau+1}\bigl(n-\tilde{k}\bigr)^{\tau-1}}
  \end{IEEEeqnarray*}
  is achievable.
\end{theorem}

We remark that the PPC scheme requires the length of each message to be $\const{L}=k\cdot \nu^{\mu}$. Note that our
proposed scheme cannot readily be obtained using the concept of refinement and lifting of so-called one-shot schemes
as introduced for PIR in \cite{DOliveiraElRouayheb18_1}, since this concept cannot readily be applied to the function
computation case.

We now provide further insight into our proposed PPC scheme by considering
the PMC scheme  as a special
case in which  the candidate set is  restricted to contain monomials.

\subsection{Special Case: PMC Scheme}
\label{sec:PIRrate_codes}

\subsubsection{Candidate Monomials}

As the rate of PMC is a decreasing function of the number of candidate monomial functions, 
we can limit ourselves to the set of monomials excluding \emph{parallel} monomials, where we define a parallel monomial
as a monomial resulting from raising another monomial to a positive integer power, i.e.,
to $\{{\vect{W}}^{\vect{i}}: \vect{i} \in (\{0\} \cup \Naturals)^{f},\, 1\leq \textsf{wt}(\vect{i})\leq g,\, \vect{i} \mid p,\, p \in \set{P}_g \}$, where $\set{P}_g$ %
 denotes the set of prime numbers less or equal to $g$ and  $\vect{i}=(i_1,\ldots,i_f) \mid p$ means that all nonzero $i_j$, $j \in [f]$, are divisors of $p$.  %
For example, for a bivariate monomial over the variables $x$ and $y$ of degree at most $g=2$ the set of possible
monomials is $\{x,y, xy, x^2,y^2\}$. Note that $x^2$ is a parallel monomial as it can be obtained by raising the
monomial $x$ to the power of $2$. Thus, $x^2$ and $y^2$ are parallel monomials and can be excluded from the set of
candidate monomials.  Denote by $\set{P}=\{ p_{1}, \ldots, p_{|\set{P}|} \}$ an arbitrary nonempty subset of $\set{P}_g$. By applying the Legendre formula for counting the
prime numbers less or equal to $g$, we obtain the 
number of nonparallel monomials as
\vspace{-0.5ex}
\begin{IEEEeqnarray*}{rCl}
	\widetilde{\const{M}}(f,g) &=& {g+f \choose g} -1   \\ \nonumber		
	 &&+\sum_{\substack{\forall \set{P} \subseteq \set{P}_g: \set{P} \neq \emptyset, \\ p_1  \cdots p_{|\set{P}|} \leq g}} (-1)^{{|\set{P}|}} \left[ {\left( \genfrac{}{}{0pt}{0}{\left\lfloor {\frac
				{g}{p_{1}\cdots p_{{|\set{P}|}}} }\right\rfloor +f}{\left\lfloor {\frac{g}{p_{1}\cdots p_{{|\set{P}|}}}
			}\right\rfloor} \right) } -1 \right], %
\end{IEEEeqnarray*}%
where $\lfloor \cdot \rfloor$ denotes the floor
function. %

We illustrate the key concept of our proposed scheme in Theorem \ref{thm:PMCrate_LagrangeCoded-DSS} with an
example. %
Note that in all examples  we assume that the \emph{index
    preparation} step has been performed to keep the desired polynomial  index private. We refer the readers to
  \cite[Sec.~IV-A]{ObeadLinRosnesKliewer18_1} for details. Before we proceed with the example, given a $\nu\times n$
PPC achievable rate matrix $\mat{\Lambda}_{\kappa,\nu}$, we define the
notion  of PPC interference matrices as follows.

\begin{definition}[{\cite[Def.~5]{ObeadLinRosnesKliewer18_1}}]
  \label{def:PPCinterference-matrices}
  For a given $\nu\times n$ PPC achievable rate matrix $\mat{\Lambda}_{\kappa,\nu}=(\lambda_{u,j})$ for
  $(\code{C},\tilde{\code{C}})$, we define the PPC interference matrices $\mat{A}_{\kappa{\times}n}=(a_{i,j})$ and
  $\mat{B}_{(\nu-\kappa){\times}n}=(b_{i,j})$ for the code $\tilde{\code{C}}$ with
  \vspace{-0.5ex}
  \begin{IEEEeqnarray*}{rCl}
    a_{i,j}& \eqdef &u \text{ if } \lambda_{u,j}=1,\,\forall j \in [n], i \in[\kappa], u \in  [\nu],\\
    b_{i,j}& \eqdef &u \text{ if } \lambda_{u,j}=0,\,\forall j \in [n], i \in[\nu-\kappa], u \in  [\nu].
  \end{IEEEeqnarray*}
\end{definition}
Note that in \cref{def:PPCinterference-matrices}, for each $j \in [n]$, distinct values of $u \in [\nu]$ should be
assigned for all $i$. Thus, the assignment is not unique in the sense that the order of the entries of each column of
$\mat{A}$ and $\mat{B}$ can be permuted.

\begin{example}
  \label{ex:PMCex_n4k2f2mu3}
  Consider two messages $\vmat{W}^{(1)}$ and $\vmat{W}^{(2)}$ that are stored in a noncolluding DSS using a $[4,2]$ RS
  code $\code{C}$. Suppose that the user wishes to obtain a monomial function $\vmat{X}^{(v)}$  from the candidate set $\{\vmat{W}^{(1)},\vmat{W}^{(2)},\vmat{W}^{(1)}\star\vmat{W}^{(2)}\}$ of
  monomial functions, i.e., $\mu=\widetilde{\const{M}}(2,2)=3$. We have
  $\tilde{k}=g(k-1)+1=3$ and 
  \vspace{-1.5ex}
  \begin{IEEEeqnarray*}{rCl}
    \mat{\Lambda}_{3,4}=
    \begin{pmatrix}
      1 & 1 & 1 & 0 
      \\
      1 & 1 & 0 & 1 
      \\
      1 & 0 & 1 & 1
      \\
      0 & 1 & 1 & 1
    \end{pmatrix}
\vspace{-0.6ex}
  \end{IEEEeqnarray*}
  is a valid PPC achievable rate matrix for $(\code{C},\tilde{\code{C}})$. From $\mat{\Lambda}_{3,4}$ we further
  obtain the interference matrices %
 \vspace{-0.2ex} 
  \begin{IEEEeqnarray*}{c}
    \mat{A}_{3\times 4} = 
    \begin{pmatrix}
      1 &1 &1 &2
      \\
      2 &2 &3 &3 
      \\
      3 &4 &4 &4 
    \end{pmatrix}
    \text{ and }
    \mat{B}_{1\times 4} = 
    \begin{pmatrix}
      4 &3 &2 &1
    \end{pmatrix}. 
    \vspace{-1ex}
  \end{IEEEeqnarray*}
  
  We simplify notation by letting $x_{t,j}=C^{(1)}_{t,j}$, $y_{t,j}=C^{(2)}_{t,j}$, and
  $z_{t,j}=C^{(1)}_{t,j}\cdot C^{(2)}_{t,j}$ for all ${t\in[\beta]}$, $j\in[n]$, where $\beta=\nu^{\mu}=64.$ Let the
  desired monomial function index be $v=1$. The construction of the query sets is briefly presented in the following
  steps.\footnote{With some abuse of notation, the generated queries are sets containing their answers, and vectors should be considered as the union of their entries.}
  
  \indent {\bf \textit{Initialization (Round ${\tau=1}$)}:} We start with ${\tau=1}$ to generate query sets for each
  database $j$ holding $\kappa^{\mu}=27$ distinct instances of $x_{t,j}$. By message symmetry this also applies to
  $y_{t,j}$ and $z_{t,j}$. 

  \indent {\bf \textit{Following Rounds ($\tau \in  [2:3]$)}:} Using the interference matrices $\mat{A}_{3\times 4}$ and
  $\mat{B}_{1\times 4}$ for the exploitation of side information for the $j$-th database, $j\in [n]$, we generate the
  desired query sets $Q^{(1)}_j(\set{D};\tau)$ by querying a number of new symbols of the desired monomial jointly
  combined with symbols from other monomials queried in the previous round from database $i\neq j$. Next, the undesired
  query sets $Q^{(1)}_j(\set{U};\tau)$ (if $\tau=2)$ are generated by enforcing message symmetry.
  We make the final modification to the query sets by removing all redundant $1$-sum types from the first
  round (see Lemma~\ref{lem:redundancy}) and update the query sets. This translates to removing the queries for $z_{t,j}$, 
  since they can be generated offline by the user given $x_{t,j}$ and $y_{t,j}$. The resulting query sets are shown in
  Table~\ref{tab:answers-table}, where $u_{a:b,j} \eqdef (u_{a,j},\ldots,u_{b,j})$ for $u =x,y,z$. The PMC rate of the scheme is equal to
  $\frac{k\nu^{\mu}}{\const{D}}=\frac{2\times4^3}{3\times4\times28}=0.3810$.
  
  \begin{table}[t]
  \centering
  \caption{Query sets for a $[4,2]$ RS-coded DSS with Lagrange encoding  storing $f=2$ messages and where the first ($v=1$) monomial is privately computed for $g=2$ and $\mu=3$.}
  \label{tab:answers-table}
  \vskip -3mm
  \Resize[\columnwidth]{
    \begin{IEEEeqnarraybox}[
      \IEEEeqnarraystrutmode
      \IEEEeqnarraystrutsizeadd{4pt}{2pt}]{v/c/v/c/v/c/v/c/v/c/v}
      \IEEEeqnarrayrulerow\\
      & j && 1 && 2 && 3 && 4\\
      \hline\hline
      & Q^{(1)}_j(\set{D};1)
      && x_{1:9,1},\, x_{10:18,1},\, x_{19:27,1} &&  x_{1:9,2},\, x_{10:18,2}, \, x_{28:36,2} && x_{1:9,3},\, x_{19:27,3},\, \,x_{28:36,3} && x_{10:18,2},\, x_{19:27,3},\,x_{28:36,4} &
      \\*\cline{1-11}      
      & Q^{(1)}_j(\set{U};1)
      && y_{1:9,1},\, y_{10:18,1},\, y_{19:27,1} &&  y_{1:9,2},\, y_{10:18,2}, \, y_{28:36,2} && y_{1:9,3},\, y_{19:27,3},\, \,y_{28:36,3} && y_{10:18,2},\, y_{19:27,3},\,y_{28:36,4} &
      \\*\cline{1-11}      
      & \multirow{6}{*}{$Q^{(1)}_j(\set{D};2)$}
      && x_{37:39,1}+y_{{\b 28:30},1} &&  x_{37:39,2}+y_{{\b 19:21},2} && x_{37:39,3}+y_{{\b 10:12},3} && x_{43:45,4}+y_{{\b 1:3},4} &
      \\ 
      & 
      && x_{40:42,1}+z_{{\b 28:30},1} &&  x_{40:42,2}+z_{{\b 19:21},2} && x_{40:42,3}+z_{{\b 10:12},3} && x_{46:48,4}+z_{{\b 1:3},4} &
      \\
      &
      && x_{43:45,1}+y_{{\b 31:33},1} &&  x_{43:45,2}+y_{{\b 22:24},2} && x_{49:51,3}+y_{{\b 13:15},3} && x_{49:51,4}+y_{{\b 4:6},4} &
      \\ 
       & 
      && x_{46:48,1}+z_{{\b 31:33},1} &&  x_{46:48,2}+z_{{\b 22:24},2} && x_{52:54,3}+z_{{\b 13:15},3} && x_{52:54,4}+z_{{\b 4:6},4} &
      \\
      &
      && x_{49:51,1}+y_{{\b 34:36},1} &&  x_{55:57,2}+y_{{\b 25:27},2} && x_{55:57,3}+y_{{\b 16:18},3} && x_{55:57,4}+y_{{\b 7:9},4} &
      \\
       & 
      && x_{52:54,1}+z_{{\b 34:36},1} &&  x_{58:60,2}+z_{{\b 25:27},2} && x_{58:60,3}+z_{{\b 16:18},3} && x_{58:60,4}+z_{{\b 7:9},4} &
      \\*\cline{1-11}      
      & \multirow{3}{*}{$Q^{(1)}_j(\set{U};2)$}
      && y_{40:42,1}+z_{37:39,1} &&   y_{40:42,2}+z_{37:39,2} &&  y_{40:42,3}+z_{37:39,3} && y_{46:48,4}+z_{43:45,4} &
      \\
      &
      && y_{46:48,1}+z_{43:45,1} &&  y_{46:48,2}+z_{43:45,2} && y_{52:54,3}+z_{49:51,3} && y_{52:54,4}+z_{49:51,4} &
      \\
      &
      && y_{52:54,1}+z_{49:51,1} &&  y_{58:60,2}+z_{55:57,2} && y_{58:60,3}+z_{55:57,3} && y_{58:60,4}+z_{55:57,4} &
      \\*\cline{1-11}      
      & \multirow{3}{*}{$Q^{(1)}_j(\set{D};3)$}
      && x_{61,1}+y_{{\r  58},1}+z_{{\r  55},1} && x_{61,2}+y_{{\r  52},2}+z_{{\r  49},2}
      && x_{61,3}+y_{{\r  46},3}+z_{{\r  43},3} && x_{62,4}+y_{{\r  40},4}+z_{{\r  37},4} &
      \\
      &
      && x_{62,1}+y_{{\r  59},1}+z_{{\r  56},1} && x_{62,2}+y_{{\r  53},2}+z_{{\r  50},2}
      && x_{63,3}+y_{{\r  47},3}+z_{{\r  44},3} && x_{63,4}+y_{{\r  41},4}+z_{{\r  38},4} &
      \\
      &
      && x_{63,1}+y_{{\r  60},1}+z_{{\r  57},1} && x_{64,2}+y_{{\r  54},2}+z_{{\r  51},2}
      && x_{64,3}+y_{{\r  48},3}+z_{{\r  45},3} && x_{64,4}+y_{{\r  42},4}+z_{{\r  39},4} &
      \\*\IEEEeqnarrayrulerow
    \end{IEEEeqnarraybox}}
\end{table}

\end{example}

\section{PPC Scheme for RS-Coded DSSs With Systematic Lagrange Encoding}
\label{sec:achievable-scheme_PMC_systematic}

In this section, we consider the case of RS-coded DSSs with systematic Lagrange encoding and first adapt the concept of
a PPC achievable rate matrix from Definition~\ref{def:codes_CstarG-nonSysPPCachievable} to this scenario by extending
\cite[Def.~14]{KumarLinRosnesGraellAmat19_1app}. In contrast to the PPC scheme in Section~III, the basic idea is to
utilize the systematic part of the RS code to recover the requested function.

\begin{definition}
  \label{def:codes_CstarG-SysPPCachievable}
  A $\nu\times n$ binary matrix $\mat{\Lambda}^{\mathsf{S}}_{\kappa,\nu}$ is called a \emph{PPC systematic achievable rate
  matrix} for $(\code{C},\tilde{\code{C}})$ if the following conditions are satisfied.
  \begin{enumerate}
  \item \label{item:3} $\mat{\Lambda}^{\mathsf{S}}_{\kappa,\nu}$ is a $\kappa$-column regular matrix, and
  \item \label{item:4} there are exactly $\kappa$ rows $\{\vect{\lambda}_i\}_{i\in[\kappa]}$ and $\nu-\kappa$ rows
    $\{\vect{\lambda}_{i+\kappa}\}_{i\in [\nu-\kappa]}$ of $\mat{\Lambda}^\mathsf{S}_{\kappa,\nu}$  such that
    $\forall\,i\in [\kappa]$, $\chi(\vect{\lambda}_i)$ contains an information set for $\tilde{\code{C}}$ and
    $\forall\,i\in [\nu-\kappa]$, $\chi(\vect{\lambda}_{i+\kappa})=[k]$.
  \end{enumerate}
\end{definition}

Using \cref{lem:redundancy}, the following theorem follows since it can be proved that a PPC systematic achievable rate matrix
$\Lambda^{\mathsf{S}}_{\kappa,\nu}$ with $(\kappa,\nu)=\bigl(k,k+\min\{k,n-\tilde{k}\}\bigr)$ always exists.

\begin{theorem}
  \label{thm:RS_PPC_rate}
 Consider a DSS that uses an $[n,k]$ RS code $\code{C}$ to store $f$ messages over $n$
  noncolluding databases using systematic Lagrange encoding. Let $\mu \in [f:\mu(f,g)]$
  be the number of candidate polynomials to be computed of degree at most
  $g$, $g(k-1)+1 \leq n$,  including the $f$ independent messages.
Then, the PPC rate
  \begin{IEEEeqnarray*}{c}
    \const{R}^\mathsf{S}_\textnormal{PPC}=\frac{\nu^{\mu}} {n\Bigl[f{k}^{\mu-1}+\sum_{\tau=2}^{\mu} {
        \rho(\mu,\tau)} k^{\mu-\tau}\bigl(\nu-k\bigr)^{\tau-1}\Bigr]},
  \end{IEEEeqnarray*}
  with $\nu = k+\min\{k,n-\tilde{k}\}$, is achievable.
\end{theorem}
\begin{table}[t]
  \centering
  \caption{Query sets for a $[4,2]$ RS-coded DSS with systematic Lagrange encoding storing $f=2$ messages and where the first ($v=1$) monomial is privately computed for $g=2$ and $\mu=3$.}
  \label{tab:answers-table2}
  \vskip -3mm
  \Resize[\columnwidth]{
    \begin{IEEEeqnarraybox}[
      \IEEEeqnarraystrutmode
      \IEEEeqnarraystrutsizeadd{4pt}{2pt}]{v/c/v/c/v/c/v/c/v/c/v}
      \IEEEeqnarrayrulerow\\
      & j && 1 && 2 && 3 && 4\\
      \hline\hline
      & Q^{(1)}_j(\set{D};1)
      && x_{1:4,1},\, x_{9:12,1} &&  x_{5:8,2},\, x_{9:12,2} && x_{1:4,3}, \,x_{5:8,3} && x_{1:4,4},\, x_{5:8,4} &
      \\*\cline{1-11}      
      & Q^{(1)}_j(\set{U};1)
      && y_{1:4,1},\, y_{9:12,1} &&  y_{5:8,2},\, y_{9:12,2} && y_{1:4,3}, \,y_{5:8,3} && y_{1:4,4},\, y_{5:8,4} &
      \\*\cline{1-11}      
      & \multirow{4}{*}{$Q^{(1)}_j(\set{D};2)$}
      && x_{13:14,1}+y_{{\b 5:6},1} &&  x_{17:18,2}+y_{{\b 1:2},2} && x_{13:14,3}+y_{{\b 9:10},3} && x_{13:14,4}+y_{{\b 9:10},4} &
      \\ 
      & 
      && x_{15:16,1}+z_{{\b 5:6},1} &&  x_{19:20,2}+z_{{\b 1:2},2} && x_{15:16,3}+z_{{\b 9:10},3} && x_{15:16,4}+z_{{\b 9:10},4} &
      \\
      &
      && x_{21:22,1}+y_{{\b 7:8},1} &&  x_{21:22,2}+y_{{\b 3:4},2} && x_{17:18,3}+y_{{\b 11:12},3} && x_{17:18,4}+y_{{\b 11:12},4} &
      \\ 
      & 
      && x_{23:24,1}+z_{{\b 7:8},1} &&  x_{23:24,2}+z_{{\b 3:4},2} && x_{19:20,3}+z_{{\b 11:12},3} && x_{19:20,4}+z_{{\b 11:12},4} &
      \\*\cline{1-11}      
      & \multirow{2}{*}{$Q^{(1)}_j(\set{U};2)$}  
      && y_{15:16,1}+z_{13:14,1} &&  y_{19:20,2}+z_{17:18,2} &&  y_{15:16,3}+z_{13:14,3}&&  y_{15:16,4}+z_{13:14,4} &
      \\
      &
      && y_{23:24,1}+z_{21:22,1} &&  y_{23:24,2}+z_{21:22,2} && y_{19:20,3}+z_{17:18,3} && y_{19:20,4}+z_{17:18,4} &
      \\*\cline{1-11}      
     & \multirow{2}{*}{$Q^{(1)}_j(\set{D};3)$}
      && x_{25,1}+y_{{\r 19},1}+z_{{\r 17},1} && x_{26,2}+y_{{\r 15},2}+z_{{\r 13},2}
      && x_{25,3}+y_{{\r 23},3}+z_{{\r 21},3} && x_{25,4}+y_{{\r 23},4}+z_{{\r 21},4} &
      \\
      &
      && x_{27,1}+y_{{\r 20},1}+z_{{\r 18},1} && x_{27,2}+y_{{\r 16},2}+z_{{\r 14},2}
      && x_{26,3}+y_{{\r 24},3}+z_{{\r 22},3} && x_{26,4}+y_{{\r 24},4}+z_{{\r 22},4} &
      \\*\IEEEeqnarrayrulerow
    \end{IEEEeqnarraybox}}
\end{table}

\begin{example}
  Consider the same scenario as in Example 1 where $n=4$, $k=2$, and $\tilde{k}=3$. It follows that $\nu=k+\min\{k,n-\tilde{k}\}=3$  
  and 
  \vspace*{-0.5ex}\begin{IEEEeqnarray*}{rCl}
    \mat{\Lambda}_{2,3}^\mathsf{S}=
    \begin{pmatrix}
      1 & 0 & 1 & 1
      \\
      0 & 1 & 1 & 1
      \\
      1 & 1 & 0 & 0
    \end{pmatrix}
  \end{IEEEeqnarray*}
  is a valid PPC systematic achievable rate matrix. We further obtain (by adapting \cref{def:PPCinterference-matrices} correspondingly)
  \begin{IEEEeqnarray*}{rCl}
    \mat{A}^\mathsf{S}_{2\times 4} = &
    \begin{pmatrix}
      1 &2 &1 &1
      \\
      3 &3 &2 &2
    \end{pmatrix}
    \text{ and } \mat{B}^\mathsf{S}_{1\times 4} = &
    \begin{pmatrix}
      2 &1 &3 &3
    \end{pmatrix}
  \end{IEEEeqnarray*}
  from $\mat{\Lambda}_{2,3}^\mathsf{S}$. The resulting query sets are shown in Table~\ref{tab:answers-table2} for $\mu=3$, where $u_{a:b,j} \eqdef (u_{a,j},\ldots,u_{b,j})$ for $u =x,y,z$, and the 
  PMC rate $\frac{k\nu^{\mu}}{\const{D}}=\frac{2\times 3^3}{2\times 4\times 15}=0.45$ is achievable.
\end{example}
	
\vspace{-0.5ex}
\section{Numerical Results}
\label{sec:results}

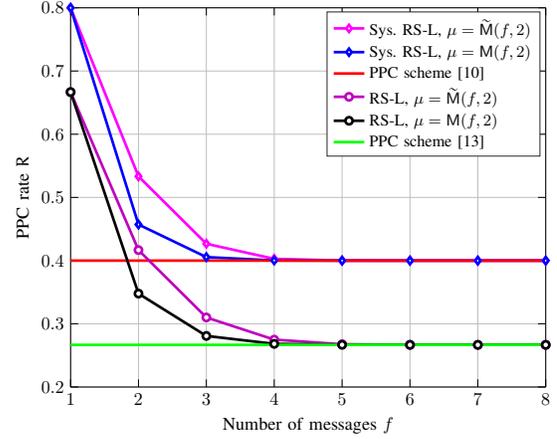
\begin{figure}[t]
	\centering
	\resizebox{0.4\textwidth}{!} 
	{
%
%
\definecolor{mycolor1}{rgb}{0.74902,0.00000,0.74902}%
\definecolor{mycolor2}{rgb}{1.00000,0.00000,1.00000}%
\begin{tikzpicture}

\begin{axis}[%
width=3.667in, 
height=2.929in, 
at={(1.95in,0.779in)},
scale only axis,
separate axis lines,
every outer x axis line/.append style={black},
every x tick label/.append style={font=\color{black}},
every x tick/.append style={black},
xmin=1,
xmax=8,
xlabel={Number of messages $f$},
every outer y axis line/.append style={black},
every y tick label/.append style={font=\color{black}},
every y tick/.append style={black},
ymin=0.2,
ymax=0.8,
ylabel={PPC rate $\const{R}$},
axis background/.style={fill=white},
xmajorgrids,
ymajorgrids,
legend style={at={(0.539,0.608)}, anchor=south west, legend cell align=left, align=left, draw=white!15!black, fill=white}
]

\addplot [color=mycolor2, line width=1.5pt, mark=diamond*, mark options={mycolor2,fill=white}]
  table[row sep=crcr]{%
1	0.8\\
2	0.533333333333333\\
3	0.426666666666667\\
4	0.40275319567355\\
5	0.400134322434899\\
6	0.400003051781096\\
7	0.400000032782557\\
8	0.400000000168802\\
};
\addlegendentry{\small Sys.~RS-L, $\mu=\widetilde{\mathsf{M}}(f,2)$}

\addplot [color=blue, line width=1.5pt, mark=diamond*, mark options={blue,fill=white}]
  table[row sep=crcr]{%
1	0.8\\
2	0.457142857142857\\
3	0.405544554455446\\
4	0.400268735112686\\
5	0.400006103608759\\
6	0.40000006556512\\
7	0.400000000337604\\
8	0.400000000000841\\
};
\addlegendentry{\small Sys.~RS-L, $\mu={\mathsf{M}}(f,2)$}

\addplot [color=red, line width=1.5pt, mark options={red,fill=white}]
  table[row sep=crcr]{%
1	0.4\\
2	0.4\\
3	0.4\\
4	0.4\\
5	0.4\\
6	0.4\\
7	0.4\\
8	0.4\\
};
\addlegendentry{\small PPC scheme \cite{Karpuk18_1}}

\addplot [color=mycolor1, mark=*,  line width=1.5pt,mark options={mycolor1,fill=white}]
  table[row sep=crcr]{%
1	0.666666666666667\\
2	0.416666666666667\\
3	0.310065982040978\\
4	0.27498016623057\\
5	0.267631411411356\\
6	0.266731030586641\\
7	0.266669123066179\\
8	0.266666720760106\\
};
\addlegendentry{\small RS-L, $\mu=\widetilde{\mathsf{M}}(f,2)$}

\addplot [color=black, mark=*,  line width=1.5pt,mark options={black,fill=white}]
  table[row sep=crcr]{%
1	0.666666666666667\\
2	0.347801892042293\\
3	0.280816587678577\\
4	0.268278462002182\\
5	0.266773957130711\\
6	0.266670760690996\\
7	0.26666675682241\\
8	0.266666667821618\\
};
\addlegendentry{\small RS-L, $\mu=\mathsf{M}(f,2)$}

\addplot [color=green, line width=1.5pt, mark options={green,fill=white}]
  table[row sep=crcr]{%
1	0.266666666666667\\
2	0.266666666666667\\
3	0.266666666666667\\
4	0.266666666666667\\
5	0.266666666666667\\
6	0.266666666666667\\
7	0.266666666666667\\
8	0.266666666666667\\
};
\addlegendentry{\small PPC scheme \cite{RavivKarpuk18_1sub}}

\end{axis}
\end{tikzpicture}%
}
	\vspace{-2ex}
	\caption{Achievable PPC rates as a function of the number of messages $f$ for $n=5$, $k=2$, and $g=2$.}
	\label{fig:PMC}
	\vspace{-1ex}
\end{figure}

In Fig.~\ref{fig:PMC}, we compare the PPC rates of Theorems~\ref{thm:PMCrate_LagrangeCoded-DSS} and
\ref{thm:RS_PPC_rate} to those of the schemes from \cite{Karpuk18_1,RavivKarpuk18_1sub} for $n=5$, $k=2$, and $g=2$. The
proposed schemes show improved performance for a low number of messages $f$. Observe that the curves converge to the
rates from \cite{Karpuk18_1,RavivKarpuk18_1sub} as the number of messages $f$ grows. In fact, it can easily be seen from
the rate expressions of Theorems~\ref{thm:PMCrate_LagrangeCoded-DSS} and \ref{thm:RS_PPC_rate} that this is always the
case (details omitted for brevity). For comparison, we also plot the PMC rate when parallell monomials are
excluded (magenta and purple lines).

\vspace{-0.5ex}
\section{Converse Bound}
Since RS codes are MDS codes and PPC can be seen as an extension of
	PLC, we can adapt the coded PLC capacity of \cite[Thm.~2]{ObeadLinRosnesKliewer18_1} to be an outer bound to the PPC
	rate. 
	However, for an infinite number of messages the PPC rates of our proposed schemes, as for the schemes of \cite{Karpuk18_1,RavivKarpuk18_1sub}, do not approach this outer bound, and it is still unknown whether the PLC capacity can be achieved by a coded PPC scheme.

\vspace{-0.03cm}


\begin{thebibliography}{10}
\vspace{-0.04cm}
\providecommand{\url}[1]{#1}
\csname url@samestyle\endcsname
\providecommand{\newblock}{\relax}
\providecommand{\bibinfo}[2]{#2}
\providecommand{\BIBentrySTDinterwordspacing}{\spaceskip=0pt\relax}
\providecommand{\BIBentryALTinterwordstretchfactor}{4}
\providecommand{\BIBentryALTinterwordspacing}{\spaceskip=\fontdimen2\font plus
\BIBentryALTinterwordstretchfactor\fontdimen3\font minus
  \fontdimen4\font\relax}
\providecommand{\BIBforeignlanguage}[2]{{%
\expandafter\ifx\csname l@#1\endcsname\relax
\typeout{** WARNING: IEEEtran.bst: No hyphenation pattern has been}%
\typeout{** loaded for the language `#1'. Using the pattern for}%
\typeout{** the default language instead.}%
\else
\language=\csname l@#1\endcsname
\fi
#2}}
\providecommand{\BIBdecl}{\relax}
\BIBdecl

\bibitem{ChorKushilevitzGoldreichSudan95_1}
B.~Chor, O.~Goldreich, E.~Kushilevitz, and M.~Sudan, ``Private information
  retrieval,'' in \emph{Proc. 36th IEEE Symp. Found. Comp. Sci.}, Milwaukee,
  WI, USA, Oct. 23--25, 1995, pp. 41--50.

\bibitem{SunJafar17_1}
H.~Sun and S.~A. Jafar, ``The capacity of private information retrieval,''
  \emph{IEEE Trans. Inf. Theory}, vol.~63, no.~7, pp. 4075--4088, Jul. 2017.

\bibitem{FreijHollantiGnilkeHollantiKarpuk17_1}
R.~Freij-Hollanti, O.~W. Gnilke, C.~Hollanti, and D.~A. Karpuk, ``Private
  information retrieval from coded databases with colluding servers,''
  \emph{SIAM J. Appl. Algebra Geom.}, vol.~1, no.~1, pp. 647--664, Nov. 2017.

\bibitem{BanawanUlukus18_1}
K.~Banawan and S.~Ulukus, ``The capacity of private information retrieval from
  coded databases,'' \emph{IEEE Trans. Inf. Theory}, vol.~64, no.~3, pp.
  1945--1956, Mar. 2018.

\bibitem{TajeddineGnilkeElRouayheb18_1}
R.~Tajeddine, O.~W. Gnilke, and S.~El~Rouayheb, ``Private information retrieval
  from {MDS} coded data in distributed storage systems,'' \emph{IEEE Trans.
  Inf. Theory}, vol.~64, no.~11, pp. 7081--7093, Nov. 2018.

\bibitem{KumarLinRosnesGraellAmat19_1app}
S.~Kumar, H.-Y. Lin, E.~Rosnes, and A.~Graell~i Amat, ``Achieving maximum
  distance separable private information retrieval capacity with linear
  codes,'' 2019, to app. in \itshape IEEE Trans. Inf. Theory\upshape. 

\bibitem{MirmohseniMaddahAli18_1}
  M.~Mirmohseni and M.~A. Maddah-Ali, ``Private function retrieval,'' in \emph{Proc. Iran Workshop Commun. Inf. Theory},  Tehran, Iran, Apr. 2018.

\bibitem{SunJafar19_1app}
H.~Sun and S.~A. Jafar, ``The capacity of private computation,'' 2019, to app.
  in \itshape IEEE Trans. Inf. Theory\upshape.

\bibitem{ChenWangJafar18_1}
Z.~Chen, Z.~Wang, and S.~Jafar, ``The asymptotic capacity of private search,''
  in \emph{Proc. IEEE Int. Symp. Inf. Theory}, Vail, CO, USA, Jun. 17--22,
  2018, pp. 2122--2126.

\bibitem{Karpuk18_1}
D.~Karpuk, ``Private computation of systematically encoded data with colluding
  servers,'' in \emph{Proc. IEEE Int. Symp. Inf. Theory}, Vail, CO, USA, Jun.
  17--22, 2018, pp. 2112--2116.

\bibitem{ObeadKliewer18_1}
S.~A. Obead and J.~Kliewer, ``Achievable rate of private function retrieval
  from {MDS} coded databases,'' in \emph{Proc. IEEE Int. Symp. Inf. Theory},
  Vail, CO, USA, Jun. 17--22, 2018, pp. 2117--2121.

\bibitem{ObeadLinRosnesKliewer18_1}
S.~A. Obead, H.-Y. Lin, E.~Rosnes, and J.~Kliewer, ``Capacity of private linear
  computation for coded databases,'' in \emph{Proc. 56th Allerton Conf.
  Commun., Control, Comput.}, Monticello, IL, USA, Oct. 2--5, 2018.

\bibitem{RavivKarpuk18_1sub}
  N.~Raviv and D.~A. Karpuk, ``Private polynomial computation from {L}agrange
  encoding,'' Dec. 2018, arXiv:1812.04142v2 [cs.IT].

\bibitem{YuLiRavivKalanSoltanolkotabiAvestimehr19_1} Q.~Yu, S.~Li, N.~Raviv, S.~M.~M. Kalan, M.~Soltanolkotabi, and
  A.~S.  Avestimehr, ``Lagrange coded computing: Optimal design for resiliency, security, and privacy,'' in
  \emph{Proc. 22nd Int. Conf. Artif. Intell. Statist.}, vol.~89, Naha, Okinawa, Japan, Apr. 16--18, 2019, pp. 1215--1225.

\bibitem{DOliveiraElRouayheb18_1}
R.~G.~L. D{'}Oliveira and S.~El~Rouayheb, ``Lifting private information
  retrieval from two to any number of messages,'' in \emph{Proc. IEEE Int.
  Symp. Inf. Theory}, Vail, CO, USA, Jun. 17--22, 2018, pp. 1744--1748.

\end{thebibliography}
\end{document}